\documentclass{jpsj2}
\title{Long-Range Magnetic Interactions Induced by the Lattice Distortions
and the Origin of the E-type
Antiferromagnetic Phase in the Undoped Orthorhombic Manganites}

\author{Igor \textsc{Solovyev}\thanks{SOLOVYEV.Igor@nims.go.jp}}

\inst{Computational Materials Science Center,
National Institute for Materials Science, 1-2-1 Sengen, Tsukuba,
Ibaraki 305-0047, Japan}

\abst{
With the increase of the lattice distortion, the orthorhombic manganites
$R$MnO$_3$ ($R$$=$ La, Pr, Nd, Tb, and Ho) are known to undergo the phase transition
from the layered A-type antiferromagnetic (AFM) state to the zigzag
E-type AFM state.
We consider the microscopic origin of this transition.
Our approach consists of the two parts.
First, we construct
an effective lattice fermion model for the manganese $3d$-bands
and derive
parameters of this model from the
first-principles electronic structure calculations.
Then, we solve this model in the Hartree-Fock
approximation (HFA) and analyze the behavior of interatomic magnetic interactions.
We argue that the nearest-neighbor interactions
decrease with the distortion and
at certain stage start to
compete
with the longer range (particularly, second- and third-neighbor)
AFM interactions in the orthorhombic ${\bf ab}$-plane, which
lead to the formation of the E-phase.
The origin of these interactions is closely related to the orbital ordering,
 which takes place in the distorted orthorhombic structure.
The model is able to capture the experimental trend
and explain why LaMnO$_3$ develops the A-type AFM order and why
it tends to transform to the E-type AFM order in
the more distorted
compounds. Nevertheless, the quantitative agreement with the experimental data
crucially depends on other factors, such as the magnetic polarization of the oxygen sites and the
correlation interactions beyond HFA.}

\kword{undoped manganites, phase diagram, magnetic interactions, first-principles calculations, effective model}

\begin{document}
\maketitle

\section{\label{sec:introduction} Introduction}

\begin{figure}[h!]
\begin{center}
\resizebox{12cm}{!}{\includegraphics{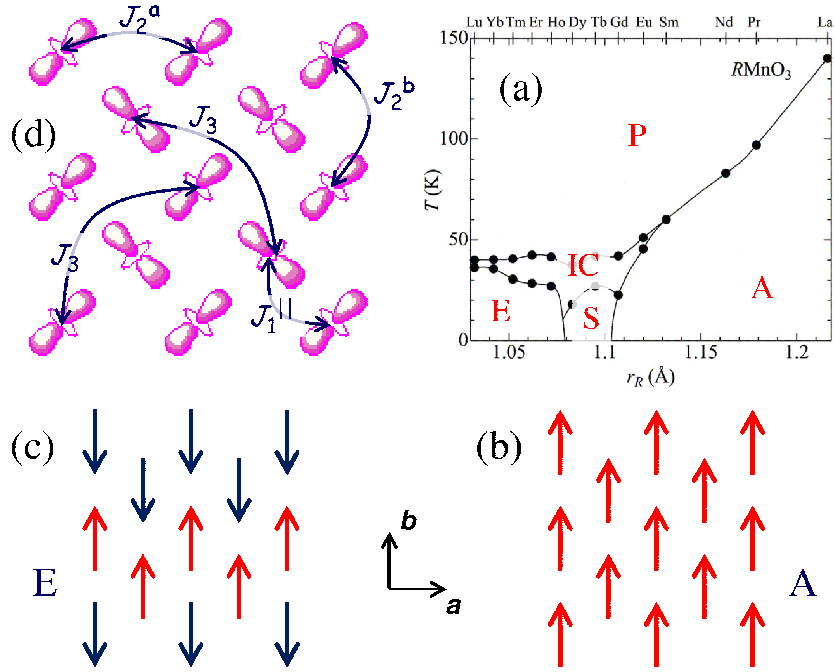}}
\end{center}
\caption{
(Color online)
(a): experimental phase diagram of
$R$MnO$_3$ versus temperature and ionic radius of rare-earth
elements (from ref. \protect\citen{Tachibana}). Magnetic phases
are denoted as paramagnetic (P), A-type AFM (A),
spiral AFM (S), incommensurate (IC), and E-type
AFM (E). (b) and (c): spin arrangement in the
orthorhombic ${\bf ab}$-plane, which takes place
in the AFM phases of the A- and E-type, respectively. (d): alternating
$3x^2$$-$$r^2$ and $3y^2$$-$$r^2$ orbitals
and main magnetic interactions
in the ${\bf ab}$-plane, which are
responsible for
the relative stability of the A- and E-states.}
\label{fig.intro}
\end{figure}

  For the long time LaMnO$_3$ was regarded as a prototypical example
of parent (or undoped) manganites, where the strong Jahn-Teller
distortion was believed to coexist with the
(layered) A-type
antiferromagnetic (AFM) state.\cite{WollanKoehler,Goodenough,Kanamori,Matsumoto,KugelKhomskii}
The origin of this AFM state was one of the most disputed points
about one decade ago, right after
the new wave of interest to the phenomenon of the colossal magnetoresistance
in the manganite
compounds has just emerged.\cite{Hamada,PickettSingh,PRL96,Sawada,Shiina,Maezono}
Despite many differences in details,
all theories of that period of time seemed to agree that the
Jahn-Teller effect plays an important role in the alternating
population of the $3x^2$$-$$r^2$ and $3y^2$$-$$r^2$ orbitals (Fig. \ref{fig.intro}),
which is primary responsible for the directional anisotropy of interatomic magnetic interactions
underlying the A-type AFM phase.
Indeed, simple considerations for the
superexchange (SE) interactions suggest that the
alternating (antiferro) ordering of the $3x^2$$-$$r^2$ and $3y^2$$-$$r^2$ orbitals
in the orthorhombic ${\bf ab}$-plane
leads to the ferromagnetic (FM or F)
coupling, while stacking (ferro) orbital ordering in the
${\bf c}$-direction is responsible for the weak AFM coupling.\cite{Goodenough2,Kanamori2,KugelKhomskii}

  The main surprise came later when it was found that after
replacing La by smaller rare-earth elements ($R$), which \textit{systematically increases}
all kinds of the
lattice
distortions (including the Jahn-Teller one), the orthorhombic $R$MnO$_3$ compounds
undergo the change of the magnetic ground state (Fig. \ref{fig.intro}).\cite{Tachibana}
Briefly, the least distorted LaMnO$_3$ forms the A-type AFM structure.
The opposite-end compounds (starting from HoMnO$_3$) form the so-called
E-type (zigzag) AFM structure. In the intermediate region, the magnetic structure is
incommensurate and keeps some features of the both A- and E-type AFM phases.
The appearance of the E-type AFM structure, which \textit{breaks the inversion symmetry}
in otherwise centrosymmetric crystal environment, is particularly interesting.
It can be hardly understood in terms of the nearest-neighbor (NN) SE
interactions alone, because such a mechanism would inevitably imply the change of
the orbital state
and
operate against the large energy gain associated with the
Jahn-Teller distortion.
Therefore, it seems that the more realistic scenario
should involve some
longer range interactions.

 At the purely phenomenological level, the competition between the A-
and E-type AFM phases in the ${\bf ab}$-plane
can be rationalized in terms of the following
interaction parameters and trends (Fig. \ref{fig.intro}):
\begin{itemize}
\item[$\bullet$]
the NN interaction $J^\parallel_1$, which, depending on its sign,
favors either FM or bipartite AFM arrangement;
\item[$\bullet$]
the 3rd-neighbor AFM interaction $J_3$
which couples all 3rd-neighbor spins antiferromagnetically, as required
for the E-type AFM structure.
Therefore,
$J_3$ should be an indispensable ingredient
of the model analysis.
As we will see below, the main details of the magnetic phase diagram of
$R$MnO$_3$ depend on the competition between $J^\parallel_1$ and $J_3$.
If considered alone,
the 3rd-neighbor AFM interactions
would
favor the formation of an infinitely degenerate group of states,
including
two zigzag AFM structures propagating along the orthorhombic
${\bf a}$- and ${\bf b}$-axes. The experimentally
observed E-type AFM structure is the one of them,
which propagates along the ${\bf a}$-axis and where the spins are
antiferromagnetically coupled along the ${\bf b}$-axis;
\item[$\bullet$]
the 2nd-neighbor AFM interactions $J^{\bf b}_2$,
which lifts the degeneracy and together with $J_3$
determines the
direction of propagation and the periodicity of the
E-type AFM phase.
The combination of $J^{\bf b}_2$ and $J_3$
appears to be sufficient to bind the directions of spins
in each of the orbital sublattices,
which are denoted as $3x^2$$-$$r^2$ or $3y^2$$-$$r^2$
in Fig. \ref{fig.intro}.
\end{itemize}
Loosely speaking, if ferromagnetic $J^\parallel_1$ dominates over $J^{\bf b}_2$ and $J_3$, the
magnetic ground state will be
of the A-type.
On the other hand, if the longer range interactions dominates,
the magnetic ground state will tend to be
of the E-type.
The last ingredient, which stabilizes the E-type AFM phase is the
small difference between the parameters $J^\parallel_1$
acting in the
FM and AFM bonds, which
can be caused by either the exchange stiction
or the orbital ordering effects.
This difference is necessary in order to stabilize the
directions of
spins in
two orbital sublattices
relative to each other.

  The purpose of this work is to show that all these features are
in fact
closely related to the crystal distortion
and the type of the orbital ordering realized in the
orthorhombic $R$MnO$_3$ compounds.
We use the same strategy as in the previous work devoted
to BiMnO$_3$.\cite{BiMnO3} First, we derive an effective low-energy
model for the Mn($3d$) bands and extract parameters of this model from
the first-principles electronic structure calculations based on the
linear-muffin-tin-orbital (LMTO) method.\cite{LMTO} Then, we solve
this model in the Hartree-Fock approximation and analyze behavior
of interatomic magnetic interactions and the total energies.

  The existence of the
long-range magnetic interactions in LaMnO$_3$
was previously considered
in ref. \citen{springer}, in the context of
the local
stability of the A-type AFM state with respect to
other magnetic states.
In the present work, we will further consolidate this idea and
argue that it constitutes the basis
for understanding the magnetic properties of
all undoped manganites.

  The paper is organized as follows. In the next two sections we briefly
discuss the main details of the experimental crystal structure
(Sec. \ref{sec:structure})
and the electronic
structure in the local-density approximation (LDA, Sec. \ref{sec:estruc}).
The construction of the model Hamiltonian for the Mn($3d$) bands
is considered in Sec. \ref{sec:model} and the strategy employed
for the analysis of this Hamiltonian is briefly reviewed in Sec. \ref{sec:dataanalysis}.
The behavior of interatomic magnetic interactions are discussed in
Sec. \ref{sec:exchange}. Sec. \ref{sec:TEnegry} is devoted to comparison with
the experimental data. Particularly, we will consider the behavior of the correlation
energies and the magnetic polarization of the oxygen sites,
which is typically missing in the
low-energy
model. Finally, the brief summary will be given in Sec. \ref{sec:summary}.

\section{\label{sec:structure} Crystal Structure}

  All considered compounds crystallize in the highly
distorted orthorhombic structure. The space group is
$D^{16}_{2h}$ in Sch\"{o}nflies notations
(No. 62 in International Tables).
The primitive cell has four formula units.
The crystal structure itself
and its implications to the magnetic properties of LaMnO$_3$
have been discussed in many details in previous
publications.\cite{Hamada,PickettSingh,PRL96}
Some crystal structure parameters are summarized in Table \ref{tab:structure}.
\begin{table}[tb]
\caption{Crystal structure parameters of $R$MnO$_3$ compounds.
$a$, $b$, and $c$ are the orthorhombic lattice constants, Mn-O
are the interatomic distances, and
$\angle$Mn-O-Mn are the bond angles (the first line is the angle in the
${\bf c}$-direction and the second line is the angle in the
${\bf ab}$-plane). All data are taken at
room temperature except for LaMnO$_3$, corresponding to 4.2 K.}
\label{tab:structure}
\begin{tabular}{lccccc}
\hline
            & LaMnO$_3$\protect\cite{Elemans} & PrMnO$_3$\protect\cite{Alonso} & NdMnO$_3$\protect\cite{Mori}
            & TbMnO$_3$\protect\cite{Blasco}  & HoMnO$_3$\protect\cite{Munoz} \\
\hline
 $a$ (\AA)                  & 5.532 & 5.449 & 5.416 & 5.302 & 5.257 \\
 $b$                        & 5.742 & 5.813 & 5.849 & 5.856 & 5.835 \\
 $c$                        & 7.668 & 7.586 & 7.543 & 7.401 & 7.361 \\
 Mn-O (\AA)                 & 1.906 & 1.909 & 1.905 & 1.889 & 1.905 \\
                            & 1.959 & 1.953 & 1.951 & 1.946 & 1.943 \\
                            & 2.188 & 2.210 & 2.227 & 2.243 & 2.222 \\
 $\angle$Mn-O-Mn ($^\circ$) & 157   & 152   & 150   & 144   & 142   \\
                            & 154   & 151   & 149   & 146   & 144   \\
\hline
\end{tabular}
\end{table}
It also includes the references to the experimental lattice parameters,
which have been used in the
calculations. Generally, the crystal distortion in $R$MnO$_3$ tends to increase in the
direction La$\rightarrow$Pr$\rightarrow$Nd$\rightarrow$Tb$\rightarrow$Ho.
For example, such a tendency is clearly seen for the $b/a$ and $b/c$ ratios as well
as for the Mn-O-Mn angles. On the other hand, the Jahn-Teller
distortion is not monotonous and takes the maximum in TbMnO$_3$.
For example,
the
ratio of the
maximal and minimal Mn-O bondlengths is $1.187$ in TbMnO$_3$
(in comparison with $1.148$ in the least distorted LaMnO$_3$), and only $1.166$
in the following it HoMnO$_3$.
This structural anomaly is directly related to the anomaly of the
crystal-field (CF) splitting, which will be discussed in Sec. \ref{sec:model}.

\section{\label{sec:estruc} Electronic Structure in the Local-Density Approximation}

\begin{figure}[h!]
\begin{center}
\resizebox{7cm}{!}{\includegraphics{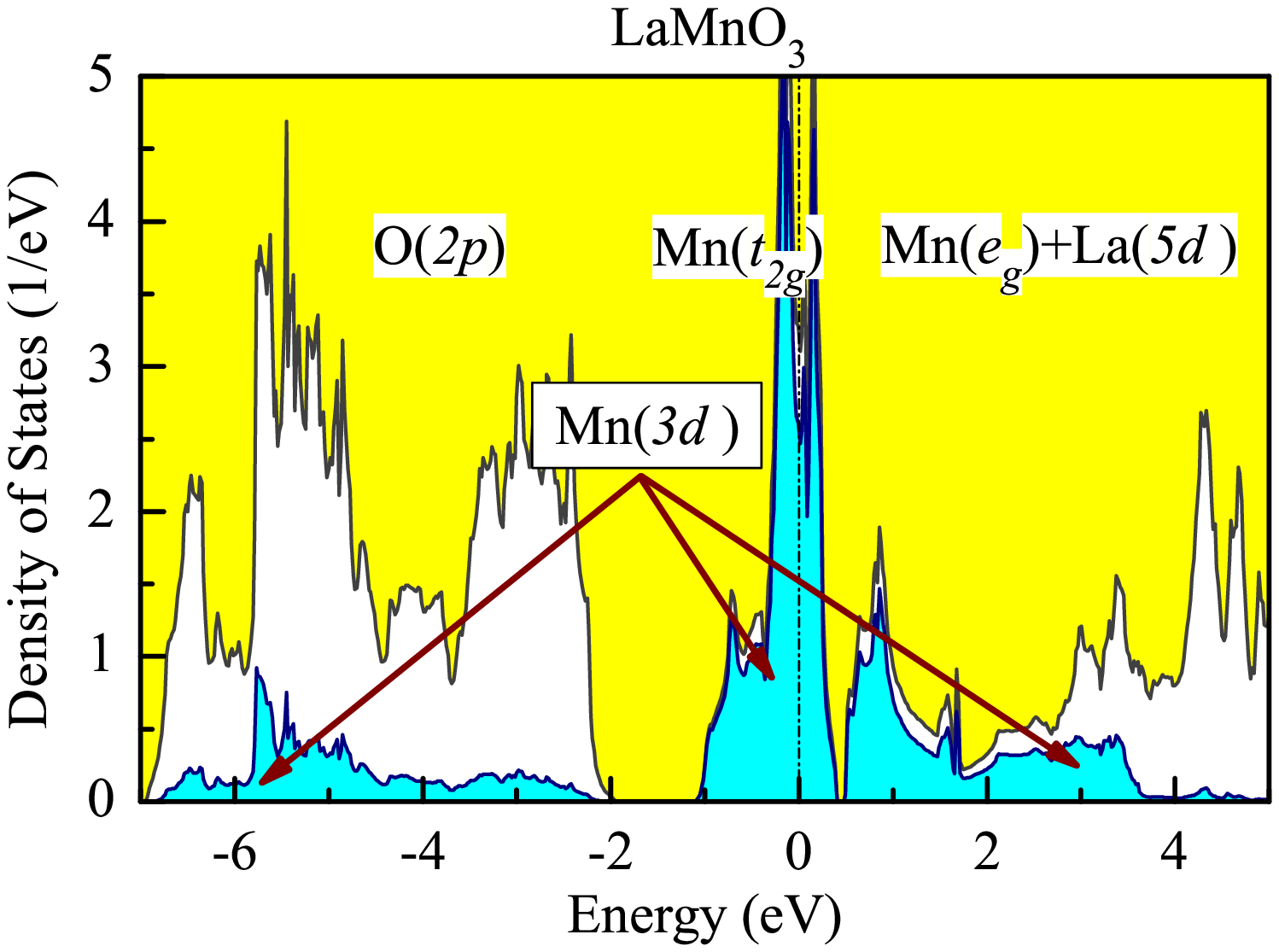}}
\resizebox{7cm}{!}{\includegraphics{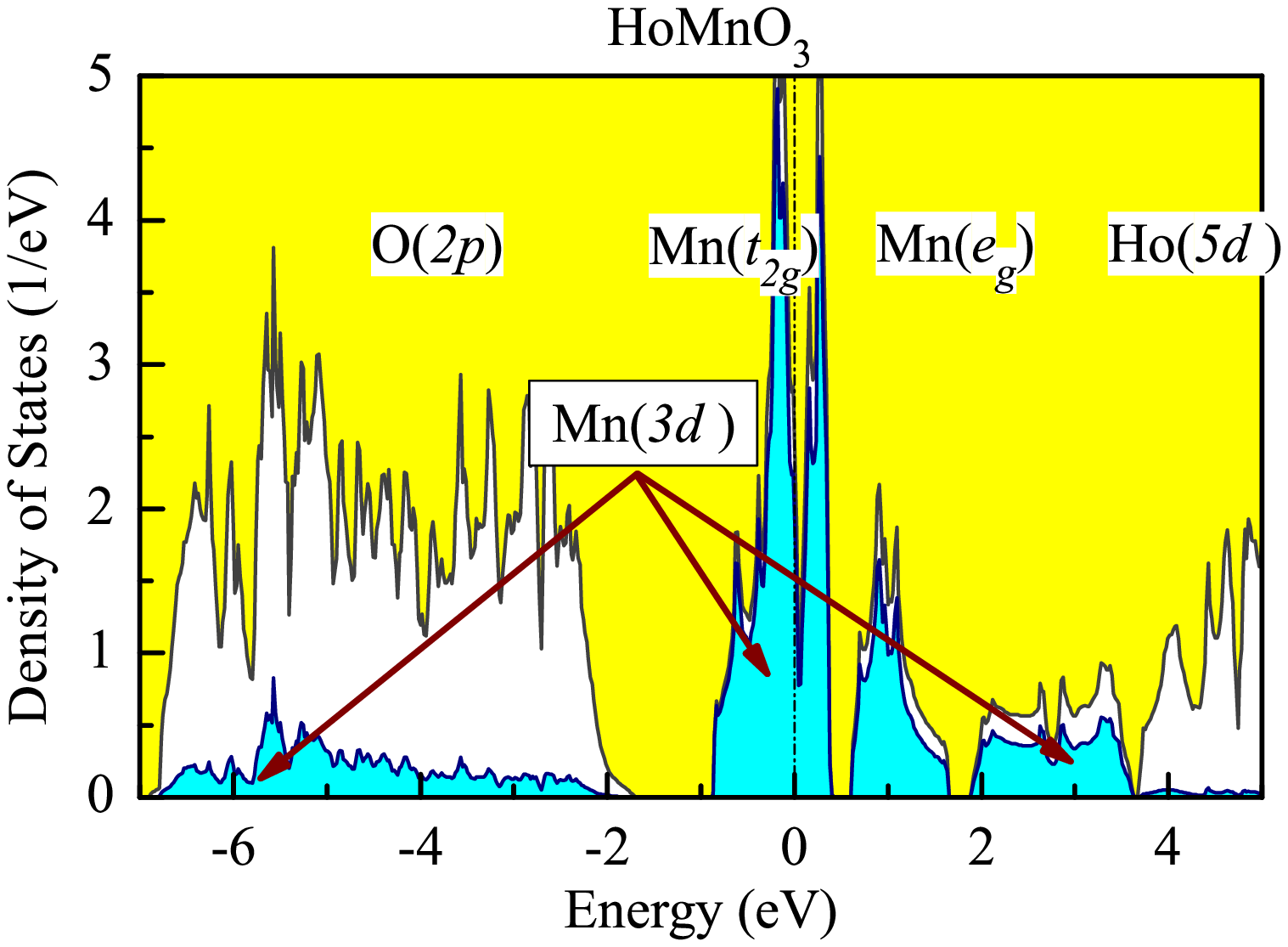}}
\end{center}
\caption{(Color online) Total and partial densities of states
as obtained
in the local-density approximation
for LaMnO$_3$ (left) and HoMnO$_3$ (right).
The shaded area shows contributions of the manganese $3d$ states.
Other symbols show the positions of the main bands.
The Fermi level is at zero energy.}
\label{fig.DOS}
\end{figure}

  An example of the LDA band structure
as obtained in the LMTO calculations
for LaMnO$_3$ and HoMnO$_3$
is shown in Fig. \ref{fig.DOS}.
The LMTO bases, which was used in the valence part of the spectrum,
typically included the Mn($3d4sp$), $R$($5d6sp$), and O($2sp$) states.
The $R$($4f$) states were treated as the (non-spin-polarized) core states.
The atomic spheres radii were determined in two steps. First, we perform the
LMTO calculations for the nominal composition, which includes 4 Mn, 4 $R$,
and 12 O atoms, and find the atomic radii from the
charge neutrality condition inside the spheres.
Then, in order to better fill the
unit cell volume and reduce the overlap between the atomic spheres,
we add 12 to 16 empty spheres with the $1s2p$-basis.
Typically, such a procedure guarantee a good agreement with the more
accurate full-potential calculations.

\begin{figure}[tb]
\begin{center}
\resizebox{!}{5cm}{\includegraphics{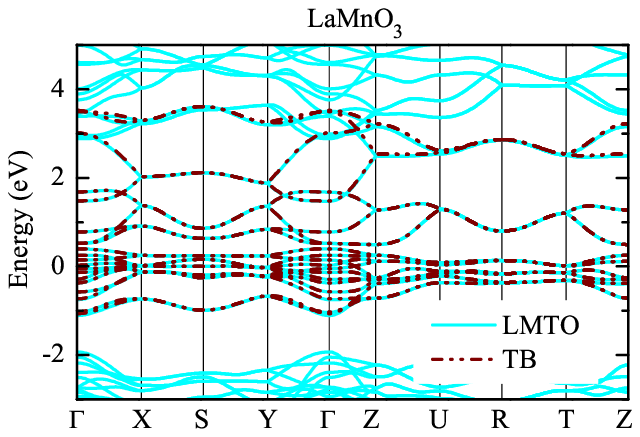}}
\resizebox{!}{5cm}{\includegraphics{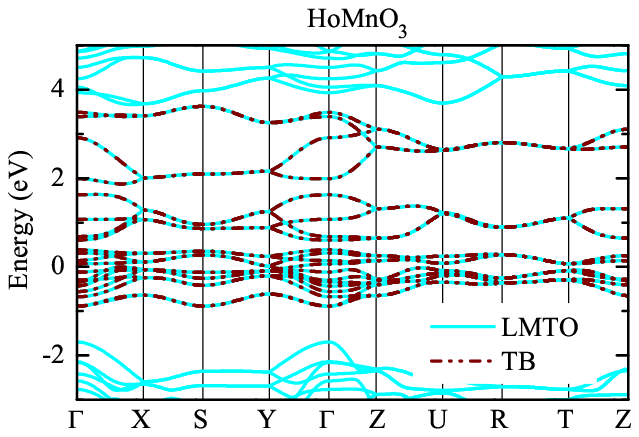}}
\end{center}
\caption{(Color online) LDA energy bands
for LaMnO$_3$ (left) and HoMnO$_3$ (right)
as obtained in the original
electronic structure calculations using the LMTO method and
after the tight-binding (TB) parametrization using the downfolding method.
Twelve low-lying bands spreading from around -1.0 till 0.4 eV
are the ``$t_{2g}$ bands'' and the next eight bands are the ``$e_g$'' bands.
Notations of the high-symmetry points of the Brillouin zone are taken
from ref. \protect\citen{BradlayCracknell}.}
\label{fig.ek}
\end{figure}

  The electronic structure near the Fermi level is mainly
formed by the
Mn($3d$)
states.
There is also a considerable weight of the
Mn($3d$) states in the oxygen band.
Due to the strong crystal-field (CF) effects in the MnO$_6$ octahedra,
the electronic structure near the Fermi level splits into the ``pseudocubic''
Mn($e_g$) and
Mn($t_{2g}$) bands.
The Jahn-Teller distortion further splits the Mn($e_g$) band
in two subbands lying at around 1 and 3 eV (Fig. \ref{fig.ek}).
In NdMnO$_3$, TbMnO$_3$, and HoMnO$_3$, these subbands are separated
by an energy gap, whereas in the least distorted LaMnO$_3$ and PrMnO$_3$,
there is a small overlap between them.
In the majority of the considered compounds, there is also a small
overlap between upper Mn($e_g$) and $R$($5d$) bands. An exception is
HoMnO$_3$, where these bands are separated by a small energy gap.

\section{\label{sec:model} Construction and Parameters of the Model Hamiltonian}

  Our next goal is the construction of an effective model Hamiltonian
for the Mn($3d$) bands located near the Fermi level.
For these purposes we use the method
proposed in ref. \citen{PRB06a}.
Many details can be found in the review article.\cite{rev08}
The model itself is specified as follows:
\begin{equation}
\hat{\cal{H}}= \sum_{{\bf R}{\bf R}'} \sum_{\alpha_1 \alpha_2} t_{{\bf
R}{\bf R}'}^{\alpha_1 \alpha_2}\hat{c}^\dagger_{{\bf
R}\alpha_1}\hat{c}^{\phantom{\dagger}}_{{\bf R}'\alpha_2} + \frac{1}{2}
\sum_{\bf R}  \sum_{ \{ \alpha \} } U^{\bf R}_{\alpha_1 \alpha_2
\alpha_3 \alpha_4} \hat{c}^\dagger_{{\bf R}\alpha_1} \hat{c}^\dagger_{{\bf
R}\alpha_3} \hat{c}^{\phantom{\dagger}}_{{\bf R}\alpha_2}
\hat{c}^{\phantom{\dagger}}_{{\bf R}\alpha_4},
\label{eqn:Hmanybody}
\end{equation}
where $\hat{c}^\dagger_{{\bf R}\alpha}$ ($\hat{c}_{{\bf R}\alpha}$)
creates (annihilates) an electron in the Wannier orbital
$\tilde{W}_{\bf R}^\alpha$ centered at the Mn-site
${\bf R}$, and $\alpha$ is
a joint index, incorporating the spin ($s$$=$ $\uparrow$ or $\downarrow$) and orbital
($m$$=$ $xy$, $yz$, $z^2$, $zx$, or $x^2$$-$$y^2$)
degrees of freedom.

  The one-electron Hamiltonian $\hat{t}_{{\bf R}{\bf R}'}$$=
$$\| t_{{\bf R}{\bf R}'}^{\alpha_1 \alpha_2} \|$
consists of the two parts:
the site-diagonal elements (${\bf R}$$=$${\bf R}'$) describe
the  crystal-field effects, whereas the off-diagonal elements
(${\bf R}$$\neq$${\bf R}'$) stand for the transfer integrals,
describing the kinetic energy of electrons.
They are derived
from the LDA band structure
by using the formal downfolding method,
which is totally equivalent to
the use of the
Wannier-basis
in the projector-operator method \cite{PRB07}.
The comparison between the original LDA bands
and the ones obtained in the
downfolding method
is shown in Fig. \ref{fig.ek}.
In LaMnO$_3$, the agreement is nearly perfect for the Mn($t_{2g}$) and the most of the Mn($e_g$) bands
located
in the low-energy part of the spectrum.
In this region, the original electronic structure
of the LMTO method
is well reproduced after the downfolding.
Since upper Mn($e_g$) bands overlap with the La($5d$) bands,
it is virtually impossible to reproduce all details of the electronic structure
in the minimal model (\ref{eqn:Hmanybody}) limited to the five Wannier-orbitals
centered at each Mn-site.
In this sense,
the electronic structure obtained in the downfolding method is only an approximation
to the original LDA band structure.
Similar situation occurs in PrMnO$_3$, NdMnO$_3$, and TbMnO$_3$.
In HoMnO$_3$, all Mn($3d$) bands are separated from the Ho($5d$) ones and
well reproduced by the downfolding method.

\begin{figure}[h!]
\centering \noindent
\resizebox{9cm}{!}{\includegraphics{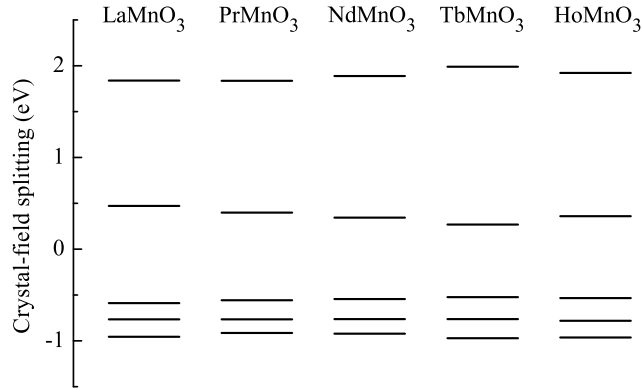}}
\caption{\label{fig.CF}
Crystal-field splitting. Three low-lying levels are of the ``$t_{2g}$''-type and the
next two levels are of the ``$e_g$''-type.}
\end{figure}

  The one-electron parameters in the real space are obtained after the
Fourier transformation.
Since we do not consider here the relativistic spin-orbit interaction,
the matrix elements
$t_{{\bf R}{\bf R}'}^{\alpha_1 \alpha_2}$ are diagonal with respect to the
spin indices: i.e.,
$t_{{\bf R}{\bf R}'}^{\alpha_1 \alpha_2}$$=$$t_{{\bf R}{\bf R}'}^{m_1 m_2} \delta_{s_1 s_2}$.
Then, the site-diagonal part of $\hat{t}_{{\bf R}{\bf R}'}$$=$$\| t_{{\bf R}{\bf R}'}^{m_1 m_2} \|$
describes the CF effects.
For example, the CF splitting is obtained after the diagonalization of
$\hat{t}_{{\bf R}{\bf R}}$.
It is particularly strong for the $e_g$ levels, being of the order of
1.5 eV  (Fig. \ref{fig.CF}), and increases with the increase of the
crystal distortion.
As was pointed out in Sec. \ref{sec:structure}, some decrease
of the $e_g$-level splitting in HoMnO$_3$ in comparison with TbMnO$_3$
is related to the decrease of the Jahn-Teller distortion.
For all considered compounds, the CF
splitting is caused by the difference in the
Mn($3d$)-O($2p$) hybridization
in different Mn-O bonds, which after the elimination of the
O($2p$)-states gives rise to the
site-diagonal elements in the model Hamiltonian. The
effect of
nonsphericity of the Madelung potential,
which plays a crucial role in the $t_{2g}$ compounds,\cite{MochizukiImada,PRB06b}
is relatively small for the $e_g$-systems.
For example in HoMnO$_3$, it changes the $e_g$-levels splitting
by less than 3\%.

\begin{figure}[h!]
\centering \noindent
\resizebox{9cm}{!}{\includegraphics{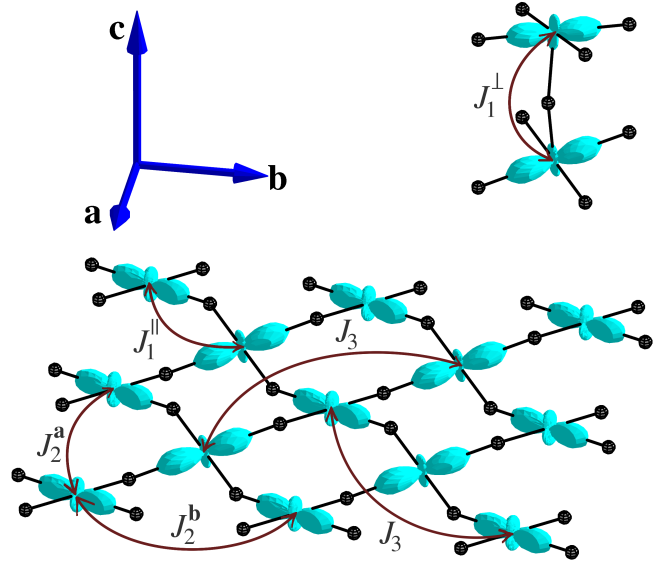}}
\caption{\label{fig.OrbitalOrdering}
Orbital ordering in LaMnO$_3$ derived from crystal-field
$e_g$ orbitals of downfolded Hamiltonian (more specifically, the
distribution of the electron density corresponding to the lowest
$e_g$ level in Fig. \protect\ref{fig.CF}).\protect\cite{remark7}
Oxygen atoms are shown by small spheres. The vectors
${\bf a}$, ${\bf b}$, and ${\bf c}$ show the directions of
orthorhombic axes. Other symbols show interatomic magnetic
interactions
in and between the planes, which are related
to the given orbital ordering.}
\end{figure}

  The directions of the CF splitting alternate on the perovskite lattice according
to the $D_{2h}^{16}$ space group.
The corresponding distribution of the $e_g$-electron densities (or the orbital ordering)
is shown in Fig. \ref{fig.OrbitalOrdering}.\cite{remark7} As will be discussed in
Sec. \ref{sec:dataanalysis}, this orbital ordering is directly responsible
for the behavior of not only the NN but also the longer range magnetic interactions.

\begin{figure}[tb]
\begin{center}
\resizebox{8cm}{!}{\includegraphics{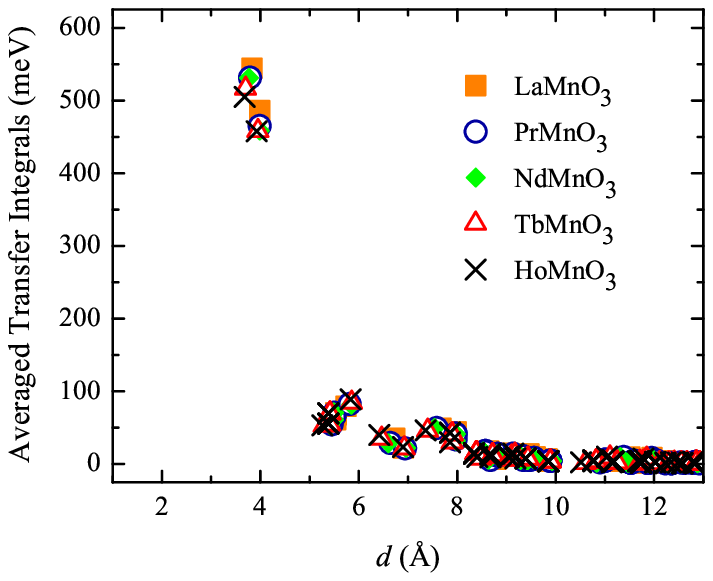}}
\end{center}
\caption{\label{fig.transfer}
(Color online)
Distance-dependence of averaged transfer integrals,
$\bar{t}_{{\bf RR}'}(d) = \left( \sum_{m_1 m_2} t^{m_1 m_2}_{{\bf RR}'}
t^{m_2 m_1}_{{\bf R}'{\bf R}} \right)^{1/2}$.}
\end{figure}

   Because of complexity of the transfer integrals,
it is rather difficult to discuss the behavior of individual
matrix elements of $\| t^{m_1 m_2}_{{\bf RR}'} \|$.
Nevertheless,
some useful information can be obtained from the analysis of
\textit{averaged} parameters
$$
\bar{t}_{{\bf RR}'}(d) = \left( \sum_{m_1 m_2} t^{m_1 m_2}_{{\bf RR}'}
t^{m_2 m_1}_{{\bf R}'{\bf R}} \right)^{1/2},
$$
where
$d$ is the distance between the Mn-sites ${\bf R}$ and ${\bf R}'$.
All transfer integrals are well localized and practically restricted by the nearest
neighbors at around 4\AA~(Fig. \ref{fig.transfer}).
Already between the next nearest neighbors,
the transfer integrals are considerably smaller.
Generally, $\bar{t}_{{\bf RR}'}$ are larger for the least distorted
LaMnO$_3$ and smaller for the more distorted HoMnO$_3$.

  The screened Coulomb interactions
$U^{\bf R}_{\alpha_1 \alpha_2 \alpha_3 \alpha_4}$
have usual dependence on the spin indices:
$U^{\bf R}_{\alpha_1 \alpha_2 \alpha_3 \alpha_4}$$=$$U^{\bf R}_{m_1 m_2 m_3 m_4} \delta_{s_1 s_2} \delta_{s_3 s_4}$.
Generally, the matrix
$\hat{U}^{\bf R}$$=$$\|  U^{\bf R}_{m_1 m_2 m_3 m_4} \|$
can
depend on the site-index ${\bf R}$.
The intersite matrix elements of $\hat{U}$ are considerably smaller.\cite{PRB06a}

  The matrix $\hat{U}^{\bf R}$ itself
has been computed in two steps \cite{PRB06a,rev08}.
First, we perform the conventional constrained LDA ($c$LDA)
calculations, and derive
parameters of on-site Coulomb
and exchange interactions between pseudoatomic Mn($3d$) orbitals.
These parameters are typically rather large
because the do not include
the so-called self-screening effects caused
\textit{by the same $3d$ electrons},
which participate in the formation of other bands
due to the hybridization \cite{rev08}.
The major contribution comes from the O($2p$) band, which has a large weight
of the Mn($3d$) states (Fig. \ref{fig.DOS}).
This channel of screening can be efficiently
taken into account in the random-phase approximation (RPA)
by starting from the interaction parameters obtained in $c$LDA
and assuming that the latter already include all other
channels of screening.\cite{PRB06a}
All RPA calculations have been performed by
starting from the LDA band structure.
Nevertheless, in order to simulate the electronic structure close
to the saturated (ferromagnetic) state, we used different Fermi levels
for the majority ($\uparrow$-) and minority ($\downarrow$-)
spin states. Namely, it was assumed that the Mn($3d$) band is
empty for the $\downarrow$-spin channel and accommodates all 16
electrons (per one primitive unit) for the $\uparrow$-spin channel.
Meanwhile, we get rid of the unphysical metallic screening
by switching off all
contributions to the RPA polarization function,
which are
associated with
the transitions within the Mn($3d$) band.\cite{rev08}

  Then, at each Mn site we obtain the $5$$\times$$5$$\times$$5$$\times$$5$
matrix $\hat{U}^{\bf R}$ of the screened Coulomb interactions.
Since the RPA screening incorporates
some effects of the local environment in solid, the symmetry of such
matrices
differs from the spherical one.\cite{rev08}
Nevertheless, just for the explanatory purposes, we fit each matrix
in terms of three parameters, which
specify interactions between the $3d$-electrons
in the spherical environment: the Coulomb repulsion $U$$=$$F^0$,
the intraatomic exchange coupling $J$$=$$(F^2$$+$$F^4)/14$, and the
``nonsphericity'' $B$$=$$(9F^2$$-$$5F^4)/441$,
where
$F^0$, $F^2$, and $F^4$ are the radial Slater's integrals.
These parameters have the following meaning:
$U$ is responsible for the charge stability of certain atomic configuration,
while $J$ and $B$ are responsible for the first and second Hund rule,
respectively. The results of such a fitting are shown in Table \ref{tab:UJB}.
\begin{table}[tb]
\caption{Results of fitting of the effective Coulomb interactions in
terms of three atomic parameters: the Coulomb repulsion $U$,
the exchange coupling $J$ and the nonsphericity $B$.
All energies are measured in eV.}
\label{tab:UJB}
\begin{tabular}{cccc}
\hline
 compound   & $U$        & $J$        & $B$     \\
\hline
 LaMnO$_3$  & $2.15$     & $0.85$     & $0.09$  \\
 PrMnO$_3$  & $2.07$     & $0.85$     & $0.09$  \\
 NdMnO$_3$  & $2.11$     & $0.85$     & $0.09$  \\
 TbMnO$_3$  & $2.24$     & $0.86$     & $0.09$  \\
 HoMnO$_3$  & $2.16$     & $0.85$     & $0.09$  \\
\hline
\end{tabular}
\end{table}
One can clearly see that the Coulomb repulsion $U$ appears to be relatively
small due to the
self-screening effects, while $J$ and $B$ are much closer to the atomic
limit.

  The model (\ref{eqn:Hmanybody}) does not explicitly include the oxygen states.
This could be a serious problem in the case of manganites, which are known
to be close to the charge-transfer regime.\cite{MizokawaFujimori}
On the other hand,
it is well know that
in many cases
a good semi-quantitative description of the magnetic properties of manganites
can be achieved already in a
minimal model
comprising only of the Mn($e_g$) bands.\cite{springer}
We will pursue the same point of view and concentrate on the
behavior of the Mn($3d$) bands. The magnetic polarization of the oxygen states
will be considered in Sec. \ref{sec:TEnegry}, where it will be also argued
that this effect is partially compensated by correlation interactions
in the Mn($3d$) band beyond the Hartree-Fock approximation.

\section{\label{sec:dataanalysis} Solution and Analysis of the Model}

  The model Hamiltonian (\ref{eqn:Hmanybody}) was solved in the
Hartree-Fock (HF) approximation.\cite{BiMnO3,PRB06b,rev08} After the solution
for each magnetic state, the total
energy changes corresponding to infinitesimal rotations of the spins
magnetic moments near this state
were mapped onto the Heisenberg model:\cite{JHeisenberg,TRN}
$$
E_{\rm Heis} = -\frac{1}{2} \sum_{{\bf RR}'} J_{{\bf RR}'} {\bf e}_{\bf R} \cdot {\bf e}_{{\bf R}'},
$$
where ${\bf e}_{\bf R}$ is the direction of the magnetic moment at the
site ${\bf R}$.
The parameters $\{ J_{{\bf RR}'} \}$ can be expressed
through the one-electron (retarded)
Green function, $\hat{\cal G}^s_{{\bf RR}'}(\omega)$, and the
spin-dependent part of the one-electron potential, $\Delta \hat{\cal V}_{\bf R}$,
obtained from the self-consistent solution of the HF equations.
For some applications, it is convenient to consider $J_{{\bf RR}'}$
as the function of the band filling:
\begin{equation}
J_{{\bf RR}'}(\omega) = \int_{-\infty}^{\omega} d \omega' \mathcal{J}_{{\bf RR}'}(\omega'),
\label{eqn:exchange}
\end{equation}
where
\begin{equation}
\mathcal{J}_{{\bf RR}'}(\omega') = \frac{1}{2 \pi} {\rm Im}
{\rm Tr}_L \left\{ \hat{\cal G}_{{\bf RR}'}^\uparrow (\omega')
\Delta \hat{\cal V}_{{\bf R}'} \hat{\cal G}_{{\bf R}'{\bf R}}^\downarrow (\omega')
\Delta \hat{\cal V}_{\bf R} \right\}
\label{eqn:integrant}
\end{equation}
and ${\rm Tr}_L$ is the trace over the orbital indices.
In order to obtain
the observable parameters,
$J_{{\bf RR}'}(\omega)$
should be taken
at the Fermi energy $\varepsilon_{\rm F}$:
$J_{{\bf RR}'} \equiv J_{{\bf RR}'}(\varepsilon_{\rm F})$.
Some details of this procedure can be found in the review article\cite{rev08}
as well as in the recent publication devoted to BiMnO$_3$.\cite{BiMnO3}

\section{\label{sec:exchange}Electronic Structure and Behavior of Interatomic Magnetic Interactions}

\begin{figure}[h!]
\centering \noindent
\resizebox{14cm}{!}{\includegraphics{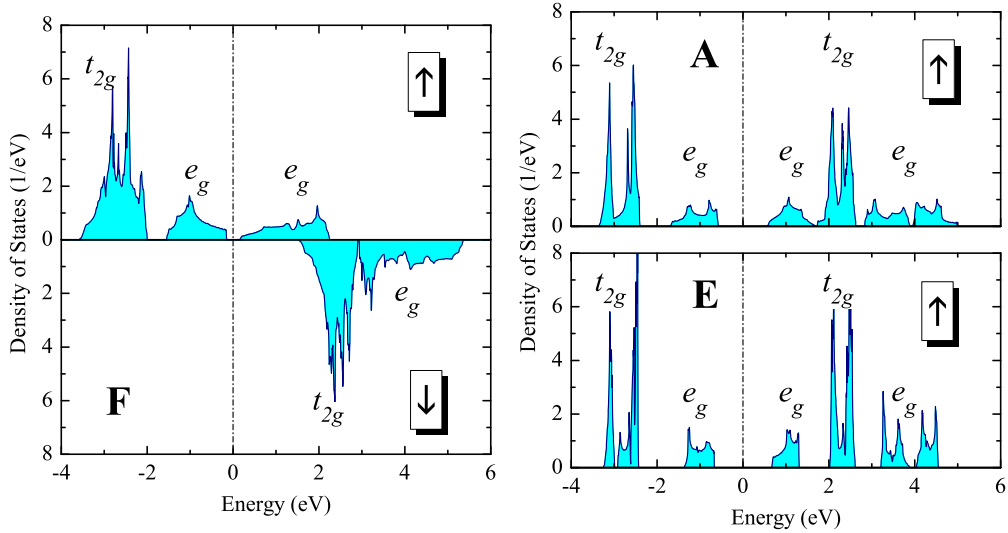}}
\caption{\label{fig.HFDOS}
(Color online)
Densities of states obtained in the Hartree-Fock
calculations for the
ferromagnetic (F),
A- and E-type
antiferromagnetic phases of
LaMnO$_3$.
The Fermi level is at zero energy
(shown by dash-dotted line).
Other symbols show the positions of the main bands.
Different spin states are indicated by the arrows.}
\end{figure}

  A typical example of the densities of states obtained in the HF calculations
for the FM and several AFM phases of LaMnO$_3$ in shown in Fig. \ref{fig.HFDOS}.
Even in LaMnO$_3$,
which is the least distorted compound, the small value of $U$,
obtained in the combined $c$LDA+RPA approach,
appears to be sufficient to open the gap in the $e_g$ band, so that all magnetic phases,
including the FM one, become insulating. As expected,
the increase of the number of the AFM bonds
associated with the change of the magnetic state
in the direction
FM$\rightarrow$A-type AFM$\rightarrow$E-type AFM
results in the narrowing of all bands.
Thus, the opening of the band gap is considerably facilitated by the
interplay of the crystal distortion with the AFM arrangement of spins. For example, even small
Jahn-Teller distorted appears to be sufficient to open the gap in the quasi-two-dimensional
FM planes of the
A-phase.\cite{JKPS,GorkovKresin}
A similar situation is expected for the quasi-one-dimensional spin chains
in the case of the E-phase.\cite{Hotta}
In other compounds,
with the increase of the crystal distortion the bandwidths will additionally decrease.
In other respects, the position of the main bands is similar to the one
displayed in Fig. \ref{fig.HFDOS}.

\begin{figure}[h!]
\centering \noindent
\resizebox{8cm}{!}{\includegraphics{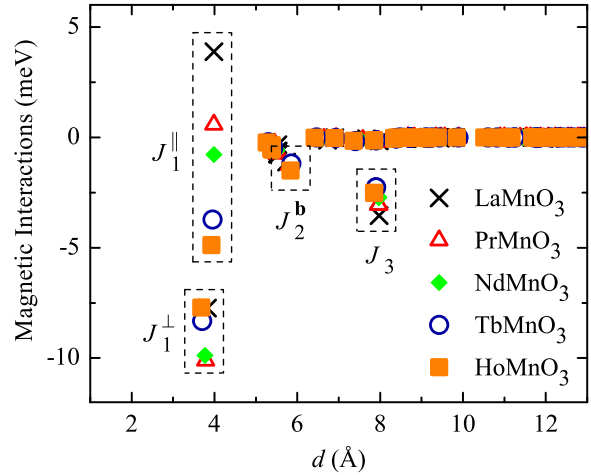}}
\caption{\label{fig.exchange1}
(Color online)
Distance-dependence of interatomic magnetic interactions,
as obtained in the Hartree-Fock calculations for the
ferromagnetic state.
The interactions, which mainly contribute to the stability
of the A- and E-type AFM phases, are shown in groups.
The notations of these interactions
are explained in Fig. \protect\ref{fig.OrbitalOrdering}.}
\end{figure}

  The distance-dependence of interatomic magnetic interactions $J_{{\bf RR}'}$
is shown in Fig. \ref{fig.exchange1}.
One can clearly distinguish four types of interactions, which
mainly contribute to the magnetic properties of $R$MnO$_3$:
the NN interaction in the orthorhombic ${\bf ab}$-plane, $J_1^\parallel$,
which strongly depends on the crystal distortion;
the NN AFM interaction along the ${\bf c}$-axis, $J_1^\perp$;
the 2nd-neighbor interaction in the ${\bf ab}$-plane, $J_2^{\bf b}$,
which operates along the orthorhombic ${\bf b}$-axis; and the 3rd-neighbor
AFM
interaction in the ${\bf ab}$-plane, $J_3$,
which operates only between those Mn-sites whose occupied
$e_g$ orbitals are pointed towards each other (see Fig. \ref{fig.OrbitalOrdering}).
Other interactions are considerably weaker.
Particularly, the 2nd-neighbor interactions along the ${\bf a}$-axis
as well as the 3rd-neighbor interactions in the direction perpendicular
to the occupied $e_g$ orbitals are small and can be neglected.

\begin{figure}[h!]
\centering \noindent
\resizebox{8cm}{!}{\includegraphics{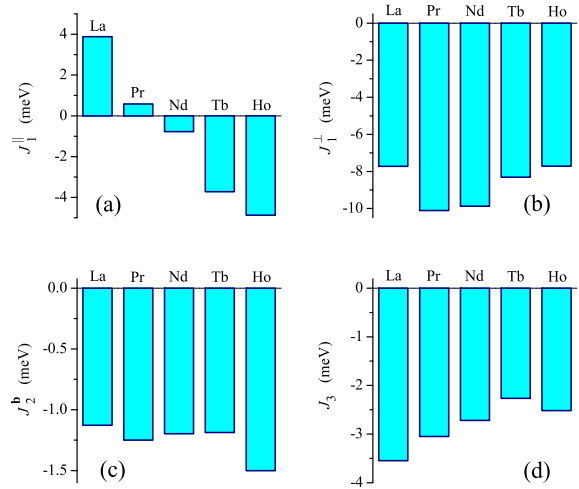}}
\caption{\label{fig.element2}
(Color online)
The behavior of the main interatomic magnetic interactions for the
$R$MnO$_3$ compounds, as obtained in the
Hartree-Fock calculations for the FM state:
the
nearest-neighbor interaction in the
${\bf ab}$-plane, $J_1^\parallel$ (a); the nearest-neighbor interaction
between the planes, $J_1^\perp$ (b); and the longer range
interactions in the ${\bf ab}$-plane, $J_2^{\bf b}$ and $J_3$
(correspondingly, b and c).
The notations of the magnetic interactions
are explained in figure \protect\ref{fig.OrbitalOrdering}.}
\end{figure}

\begin{figure}[h!]
\centering \noindent
\resizebox{6cm}{!}{\includegraphics{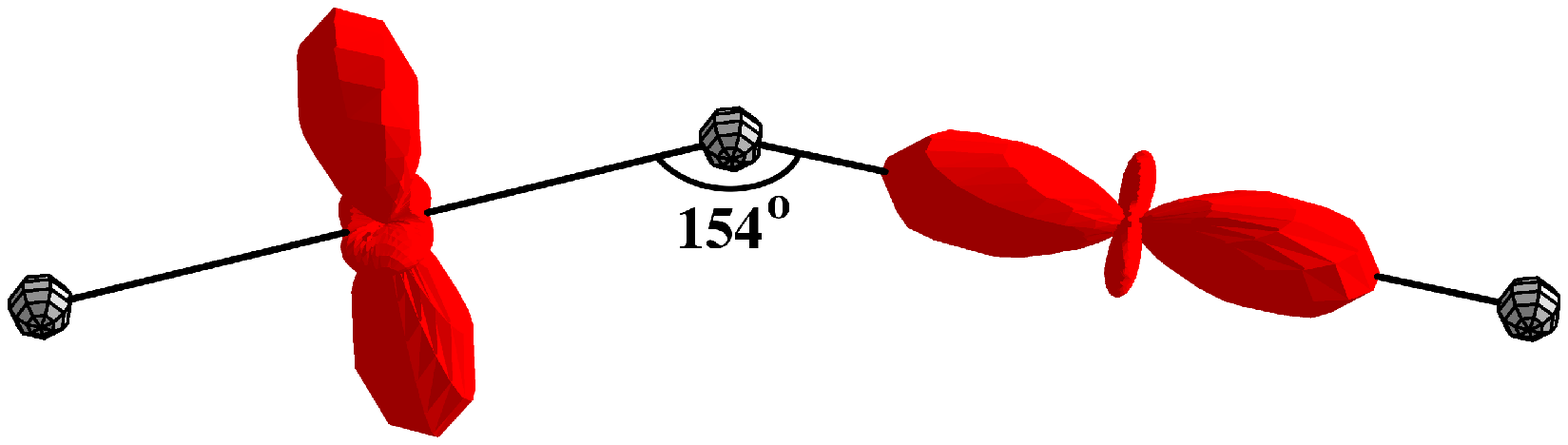}}
\resizebox{6cm}{!}{\includegraphics{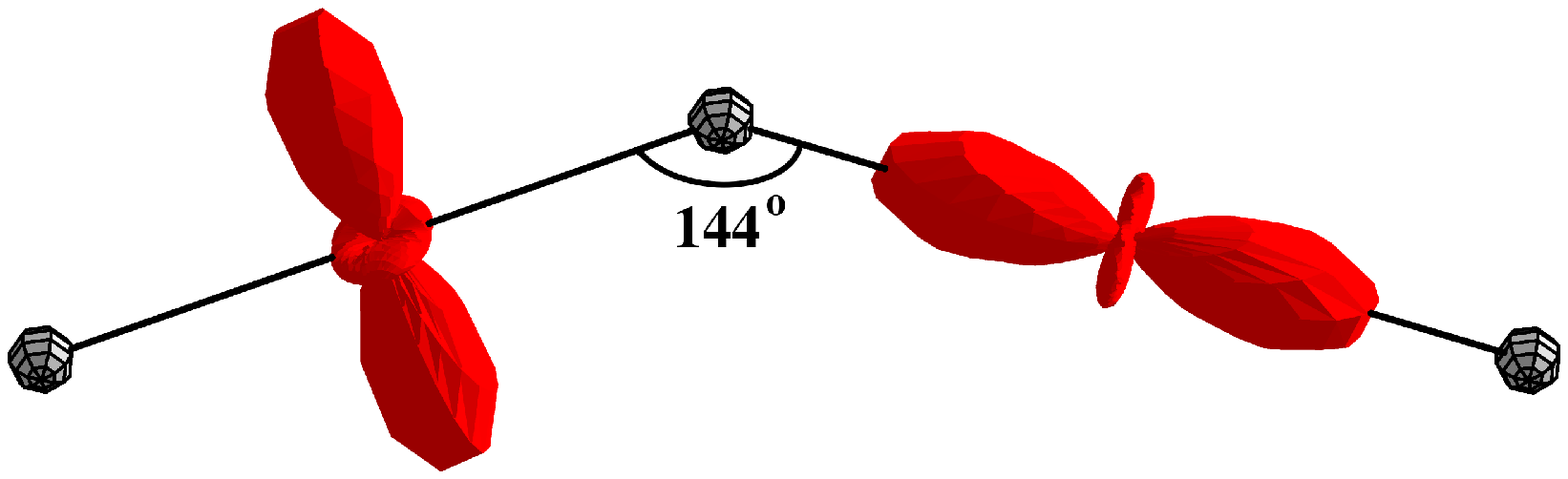}}
\caption{\label{fig.OrbitalOrderingMnOMn}
(Color online)
Fragment of the orbital ordering in the plane formed by the
single
Mn-O-Mn bond in the case of LaMnO$_3$ (left) and HoMnO$_3$ (right).}
\end{figure}

  The details of the behavior of the main magnetic interactions are shown in Fig. \ref{fig.element2}.
The interaction
$J_1^\parallel$ appears to be the most affected by the crystal distortion.
When the crystal distortion increases in the direction
La$\rightarrow$Pr$\rightarrow$Nd$\rightarrow$Tb$\rightarrow$Ho,
$J_1^\parallel$ gradually decreases and changes the sign at around Pr-Nd.
Thus, the NN coupling in the ${\bf ab}$-plane is FM at the
beginning of the series and becomes AFM at the end of it.
At the phenomenological level, such a behavior can be related to
the change of the orbital ordering in the Mn-O-Mn bond (Fig. \ref{fig.OrbitalOrderingMnOMn}).
In LaMnO$_3$, the Mn-O-Mn angle is closer to 180$^\circ$ (Table \ref{tab:structure}).
Therefore, the arrangement
of the occupied $e_g$-orbitals at the neighboring Mn-sites is nearly
``antiferromagnetic'',\cite{remark1} which according to the Goodenough-Kanamori rules
should correspond to the the FM coupling between the spins.\cite{Goodenough2,Kanamori2,KugelKhomskii}
In HoMnO$_3$, the deviation of the
Mn-O-Mn angle from 180$^\circ$ is substantially larger. Therefore, the
``antiferromagnetic orbital ordering'' is strongly distorted so that the
spin coupling can become AFM.
Nevertheless,
as we will see below, although such a phenomenological
interpretation is strongly affected by other
details
of the electronic structure
and particularly -- by the hybridization between the $t_{2g}$ and $e_g$ states, which
is caused by the crystal distortion.

  Other magnetic interactions also depend on the
crystal distortion. However, the distortion does not
change the character of these interactions, and $J_1^\perp$, $J_2^{\bf b}$ and $J_3$
are AFM for all considered compounds.

  The most striking result of the present calculations is the
existence of relatively strong
longer range
AFM interactions $J_2^{\bf b}$ and $J_3$.
The appearance of $J_3$ is expected
for the given type of the orbital ordering (Figs. \ref{fig.intro} and \ref{fig.OrbitalOrdering}).
It operates between
such
3rd neighbor sites
${\bf R}$ and ${\bf R}'$
in the ${\bf ab}$-plane,
whose occupied $e_g$ orbitals are directed
towards each other, and is mediated by the intermediate site, whose occupied
$e_g$ orbital is nearly orthogonal to the bond $\langle {\bf RR}' \rangle$.
Although the direct transfer integrals
between such sites ${\bf R}$ and ${\bf R}'$ are
small (Fig. \ref{fig.transfer}, note that the distance between
3rd neighbors in the ${\bf ab}$-plane is about 8 \AA),
the on-site Coulomb repulsion $U$ is also
relatively small (Table \ref{tab:UJB}).
Therefore, the longer range AFM interactions,
which are mediated by unoccupied $e_g$ orbitals of intermediate Mn-sites,
have the same origin as the SE interactions,
operating in the
charge-transfer insulators via the oxygen states \cite{Oguchi,ZaanenSawatzky,PRB98}, and the mechanism itself
can be called the ``super-superexchange''.
Another 3rd-neighbor interaction, operating between Mn-sites in the ${\bf ab}$-plane
whose occupied
$e_g$ orbital are nearly orthogonal to the bond connecting these sites, is negligibly small.
A similar situation occurs in the low-temperature monoclinic phase of BiMnO$_3$.\cite{BiMnO3}
The main difference is that
the orbital ordering realized in
BiMnO$_3$ is different from the one which takes place in the
orthorhombic compounds. Therefore, the
long-range AFM interactions in BiMnO$_3$
will tend to stabilize another magnetic state, which
is also different from the E-state.

  The mechanism responsible for the appearance of
the relatively strong interaction $J_2^{\bf b}$ is not so straightforward.
Nevertheless, as we will show below, some useful information can be gained
from the analysis of the
band-filling dependence of the 2nd-neighbor interactions in the ${\bf ab}$-plane.

\begin{figure}[h!]
\centering \noindent
\resizebox{12cm}{!}{\includegraphics{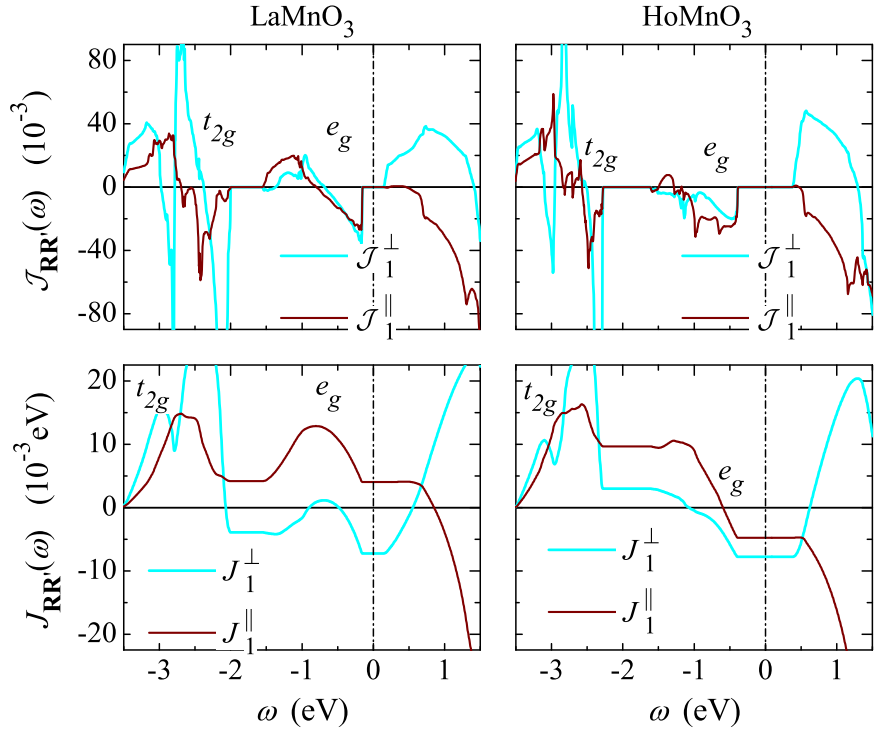}}
\caption{\label{fig.Jband1}
(Color online)
Band-filling dependance of the nearest-neighbor
magnetic interactions in the ${\bf ab}$-plane ($J^\parallel_1$) and
between the planes ($J^\perp_1$).
The magnetic interactions were calculated in the FM state for
LaMnO$_3$ (left) and HoMnO$_3$ (right).
Upper panel shows the behavior of the integrant (\protect\ref{eqn:integrant}),
while the lower panel shows the exchange coupling (\protect\ref{eqn:exchange}).
The Fermi level is at zero energy
(shown by dash-dotted line).
The positions of the $t_{2g}$- and $e_g$-bands are indicated by symbols.}
\end{figure}

  Fig. \ref{fig.Jband1} shows the behavior of the NN magnetic interactions
as a function of the band filling.
Somewhat unexpectedly,
the NN interactions in LaMnO$_3$ are mainly formed by the $t_{2g}$-band. Particularly,
the values of
both
$J^\parallel_1$ and $J^\perp_1$
are well reproduced already after integration over the
$t_{2g}$-band spreading from -3.5 eV till -2.0 eV.
The distribution of $\mathcal{J}_{{\bf RR}'}$
in the region of
the occupied $e_g$-band is
antisymmetric. Therefore, there is a strong cancelation of contributions
to $J_{{\bf RR}'}$ coming from
the bottom and the top of the occupied $e_g$-band,
so that the total integral (\ref{eqn:exchange}) over the $e_g$-band practically vanishes.
In this sense, our explanation for the
A-type AFM order in LaMnO$_3$
is rather different from the one
adopted in the model calculations,\cite{PRL96,Maezono,Shiina}
which typically do not consider the rotations of the MnO$_6$ octahedra.
According to the present calculations, the behavior of the
NN magnetic interactions in LaMnO$_3$ is mainly related to
the hybridization between the atomic $t_{2g}$- and $e_g$-orbitals,
which is induced by these rotations.
Without the hybridization, all contribution of the half-filled $t_{2g}$-band
to the NN magnetic interactions are expected to be antiferromagnetic.\cite{Maezono,Shiina}
Our analysis shows that the hybridization can easily change
the character of these interactions.

  The $t_{2g}$-$e_g$ hybridization becomes even stronger in the more distorted HoMnO$_3$,
so that the contributions of the $t_{2g}$-band become \textit{ferromagnetic}
both for $J^\parallel_1$ and $J^\perp_1$.
On the contrary,
all contributions of the $e_g$-band to the NN interactions are antiferromagnetic.
Therefore,
the $e_g$-band is totally responsible for the AFM character of
NN magnetic interactions in the case HoMnO$_3$.

\begin{figure}[h!]
\centering \noindent
\resizebox{12cm}{!}{\includegraphics{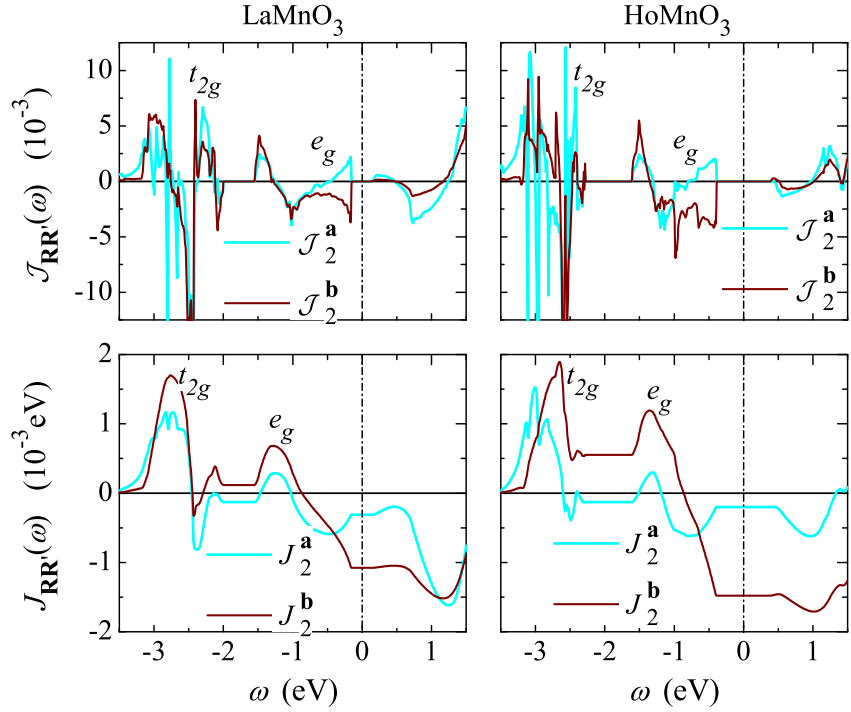}}
\caption{\label{fig.Jband2}
(Color online)
Band-filling dependance of the
second-neighbor
magnetic interactions in
the ${\bf ab}$-plane.
The magnetic interactions were calculated in the FM state for
LaMnO$_3$ (left) and HoMnO$_3$ (right).
Upper panel shows the behavior of
the integrant (\protect\ref{eqn:integrant}),
while the lower panel shows the exchange coupling (\protect\ref{eqn:exchange}).
The notations of the magnetic interactions are explained in Fig. \protect\ref{fig.OrbitalOrdering}.
The Fermi level is at zero energy
(shown by dash-dotted line).
The positions of the $t_{2g}$- and $e_g$-bands are indicated by symbols.}
\end{figure}

  The behavior of 2nd-neighbor interactions in the ${\bf ab}$-plane as a
function of the band filling is shown in Fig. \ref{fig.Jband2}.
Generally, the integrant $\mathcal{J}_{{\bf RR}'}(\omega)$ oscillates in sign.
Moreover,
as the distance between the lattice centers ${\bf R}$ and ${\bf R}'$ increases,
the number of such oscillations also increases.
This property can be rigorously proven for the
tight-binding bands, assuming that all transfer integrals
(or ``hoppings'')
are restricted by the nearest neighbors.
Then, the number of nodes of $\mathcal{J}_{{\bf RR}'}(\omega)$
becomes proportional to the minimal number of hopes, which are
required in order to reach the center ${\bf R}'$
starting from the center ${\bf R}$.\cite{Heine1,Heine2}
Thus, $\mathcal{J}_{{\bf RR}'}(\omega)$ is expected to have more nodes
for the 2nd-neighbor interactions in comparison with the NN ones,
as it is clearly seen from the comparison of Figs. \ref{fig.Jband1} and \ref{fig.Jband2}.
Nevertheless, the lattice distortion and orbital ordering effects can cause some
violation of these simple tight-binding rules.
Let us consider the behavior of $\mathcal{J}_{{\bf RR}'}(\omega)$ in the region
of the $e_g$-band, where $\mathcal{J}^\parallel_1(\omega)$ has only one node,
which is qualitatively consistent with the tight-binding rules.
Then,
$\mathcal{J}^{\bf a}_2(\omega)$ has two nodes, which is again consistent with the tight-binding rules.
Such a behavior is responsible for the
strong cancelation of positive and negative contributions to $J^{\bf a}_2$
in the process of integration over $\omega$ and readily explains the fact that
the final values of $J^{\bf a}_2$ are relatively small for all considered compounds.
However, the $\omega$-dependence of $\mathcal{J}^{\bf b}_2$ appears to be strongly deformed.
In the region of the $e_g$-band it has only one node . Therefore, the strong cancelation,
which took place for $J^{\bf a}_2$, does not occurs for $J^{\bf b}_2$.
This leads to the strong anisotropy of the 2nd-neighbor
interactions in the ${\bf ab}$-plane, $|J^{\bf b}_2| \gg |J^{\bf a}_2|$,
which plays a vital role in the formation of the E-type AFM structure.
Particularly, it readily explains the fact why
the FM zigzag chains in the observed E-type AFM structure propagate
along the ${\bf a}$-direction and are antiferromagnetically
coupled along the ${\bf b}$-axis (and not vice versa).

  Thus, the behavior of the main magnetic interactions
replicates the gradual change of the crystal distortion.
The form of both NN and long-range magnetic interactions is
closely
related to the orbital ordering
realized in the distorted orthorhombic structure.
Particularly, the crystal distortion explains
\begin{itemize}
\item[$\bullet$]
the gradual change of $J^\parallel_1$ from FM
in the case of LaMnO$_3$ to AFM at the end of the series.
Near the point of the FM-AFM crossover, $J^\parallel_1$ is small
and the magnetic ground state is mainly controlled by the
longer range interactions.
\item[$\bullet$]
the existence of the longer range AFM interactions $J_2^{\bf b}$ and $J_3$,
which
bind the spin magnetic moments within each orbital sublattice, and
determine both the direction of propagation and the periodicity of the
E-phase.
\end{itemize}

  Nevertheless, there should be an additional mechanism responsible for the
relative orientation
of spin magnetic moments in two orbital sublattices, which are marked as
$3x^2$$-$$r^2$ and $3y^2$$-$$r^2$ in Fig. \ref{fig.intro}.
Since each
spin in the E-type AFM structure
participates in the formation of two FM and two AFM bonds with the
nearest neighbors in the ${\bf ab}$-plane,
some difference between parameters $J^\parallel_1$
acting in the FM and AFM bonds
is required in order to fix the directions of spins
in the two orbitals sublattices relative to each other.\cite{remark8}
Such a modulation of the parameters $J^\parallel_1$ can be caused
by several mechanisms.
Generally, once the symmetry is broken by the AFM spin order,
orbital and lattice degrees of freedom
will tend to adjust this symmetry change.

\begin{figure}[h!]
\centering \noindent
\resizebox{10cm}{!}{\includegraphics{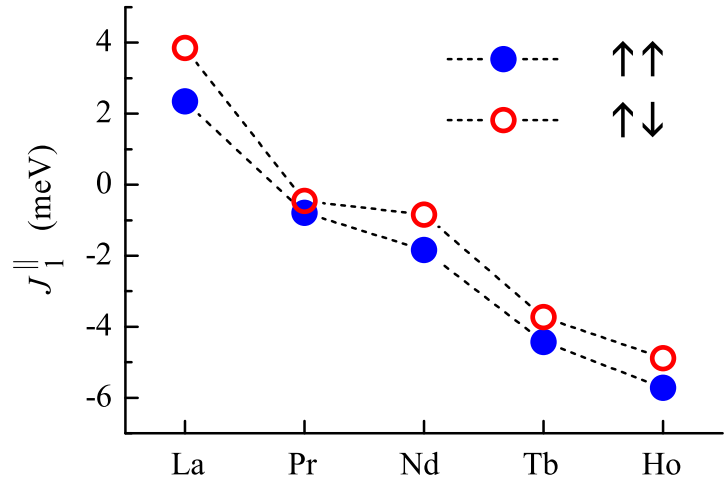}}
\caption{\label{fig.J1an}
(Color online)
Nearest-neighbor magnetic interactions in the
${\bf ab}$-plane of E-type antiferromagnetic phase.
The magnetic coupling in the FM and AFM
bonds is denoted as
$\uparrow \uparrow$ and $\uparrow \downarrow$, respectively.}
\end{figure}

  One mechanism is purely
electronic and related to the
small deformation of the orbital ordering in the AFM phase.
For example, in BiMnO$_3$ such a mechanism facilitates the
formation of the $\uparrow \downarrow \downarrow \uparrow$ AFM structure, which breaks
the inversion symmetry.\cite{BiMnO3} Nevertheless, in $R$MnO$_3$ the situation appears to be different.
For all considered compounds, the NN interactions calculated in the
E-phase satisfy the following condition:
$J^\parallel_1 (\uparrow \uparrow) < J^\parallel_1 (\uparrow \downarrow)$,
where the notations
$\uparrow \uparrow$ and $\uparrow \downarrow$ are referred to
the FM and AFM bonds, respectively (Fig. \ref{fig.J1an}).
Thus, as far as the NN interactions are concerned, the E-phase appears to be unstable
with respect to the spin rotations of two orbital sublattices relative to each other.\cite{remark2}
Apparently, such a situation is realized in the intermediate region,
corresponding to the
IC- and S-states in Fig. \ref{fig.intro}.
Nevertheless, in order to stabilize the E-phase, we need another mechanism,
which enforces the inequality $J^\parallel_1 (\uparrow \uparrow) > J^\parallel_1 (\uparrow \downarrow)$.
Such a mechanism does exist and is related to the atomic displacements,
which further minimize the total energy of the system
via magneto-elastic interactions.\cite{Wang,Picozzi07}
Although we do not consider it in the present work, from rather general
properties of the double exchange and SE interactions,\cite{remark5}
it is reasonable to expect that
the AFM character of $J^\parallel_1 (\uparrow \downarrow)$ can be
enforced by the conditions, which
further
\textit{enhance} of the transfer integrals in the AFM bond.\cite{PRL99}
This can be achieved by either \textit{shrinking} the Mn-Mn bond or
\textit{increasing} the Mn-O-Mn angle.
The opposite distortions will favor the FM coupling, which are relevant to $J^\parallel_1 (\uparrow \uparrow)$.

\section{\label{sec:TEnegry}Total Energies and Comparison with the
Experimental Data}

  In this section we consider the quantitative aspects of the problem.
Particularly, we investigate whether
the the experimental phase shown in Fig. \ref{fig.intro}
can be reproduced by
the low-energy model (\ref{eqn:Hmanybody})
for the Mn($3d$) bands
and, if not, which ingredients
are missing in the model.

\begin{figure}[h!]
\centering \noindent
\resizebox{14cm}{!}{\includegraphics{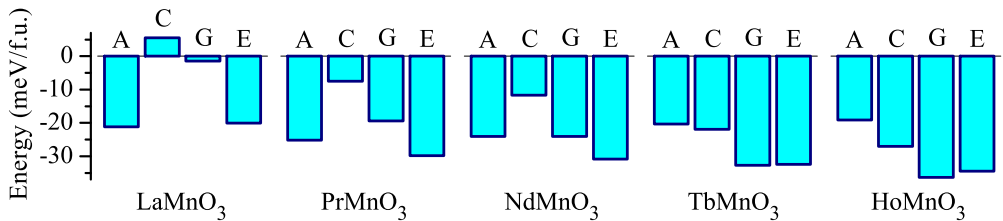}}
\caption{\label{fig.TotalE}
(Color online)
Total energies of different AFM states
obtained
for the model (\protect\ref{eqn:Hmanybody})
in the Hartree-Fock approximation.
All energies are measured relative to the FM states.
The notations of the AFM states are standard for the manganites
(see, for example, ref. \protect\citen{Hamada}).
Apart from the A- and E-states, the C-state corresponds to
the FM chains propagating along the ${\bf c}$-axis, which are antiferromagnetically
coupled in the ${\bf ab}$-plane, and the G-state corresponds to the
AFM coupling between all six nearest neighbors.}
\end{figure}

  We begin with the total energy calculations
for the model (\ref{eqn:Hmanybody}) in the HF approximation (\ref{fig.TotalE}).
In LaMnO$_3$, the lowest energy corresponds to the A-type AFM state, in agreement
with the experiment. However, the next E-type AFM state is
separated from the A-state by only
1.1 meV per one formula unit.
In PrMnO$_3$ and NdMnO$_3$,
the energy of
the E-type AFM state
appears to be lower than the one of the A-state, although experimentally
both of these compounds are the A-type antiferromagnets (Fig. \ref{fig.intro}).
Finally, for TbMnO$_3$ and HoMnO$_3$, the
model (\ref{eqn:Hmanybody}) yields the G-type
AFM ground state, where all NN spins are coupled antiferromagnetically.
Thus, although the model (\ref{eqn:Hmanybody}) predicts the
change of the magnetic ground state,
it clearly overestimates the tendencies towards
the antiferromagnetism, so that the transition
from the A- to E-type AFM state
is expected in the wrong place
(around PrMnO$_3$ and NdMnO$_3$ instead of HoMnO$_3$). The correlation interactions
beyond the HF approximation, will additionally stabilize the AFM states,\cite{PRB06b} and
only worsen the agreement with the experimental data. Therefore, before considering the
correlation effects, one should find some mechanism, which works in the
opposite direction and
additionally stabilizes the FM interactions.

\begin{figure}[h!]
\centering \noindent
\resizebox{7cm}{!}{\includegraphics{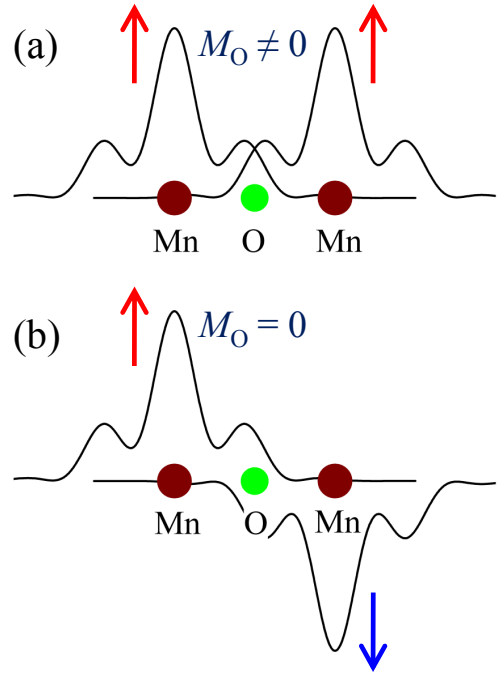}}
\caption{\label{fig.Cartoon}
(Color online)
Polarization of the
oxygen sites caused by the tails of the Wannier functions
centered at the manganese sites. In the perovskite structure,
each oxygen site is located near the midpoint between two
manganese sites. Then, in the case of the FM alignment,
the tails from the Mn-sites have the same direction of spins,
yielding the net magnetic moment also at the oxygen sites.
In the case of the AFM arrangement, these
tails cancel each other and the oxygen atoms
remains nonmagnetic.}
\end{figure}

  Such a mechanism can be related to
the magnetic polarization of the oxygen sites.\cite{WeiKu,Mazurenko}
Although the model (\ref{eqn:Hmanybody}) is designed for the
Mn($3d$) bands, the Wannier functions, which constitute the basis of the
low-energy
model (\ref{eqn:Hmanybody}), may have some tails spreading to the
oxygen and other atomic sites. The weight of these tails
in the Wannier functions
is proportional
to the weight of the O($2p$)-states in the total density of states
for the Mn($3d$) bands (Fig. \ref{fig.DOS}). In the case of the FM alignment
of the Mn-spins,
these tails will lead to some finite polarization at the
intermediate oxygen sites (Fig. \ref{fig.Cartoon}).
Since the intraatomic exchange coupling $J_{\rm O}$
associated with the oxygen atoms
is exceptionally large,\cite{Mazurenko,MazinSingh,NJP08}
even small polarization can lead to
the substantial energy gain.
This contribution is missing in the model (\ref{eqn:Hmanybody}),
where the form of the Coulomb and exchange interactions is assumed to be the same as in the
limit of isolated Mn-atoms.
In the case of the AFM alignment, the tails of the Wannier functions cancel each
other and the net magnetic polarization at the oxygen sites is zero.

  Below we present quantitative estimates of this effect for HoMnO$_3$.
By
expanding the Wannier functions over the original LMTO basis functions,\cite{PRB06a,PRB06b} one can
find the distribution of the magnetic moments over all sites of the perovskite lattice
in different magnetic structures.
The obtained values of the magnetic moments at the oxygen sites, $M_{\rm O}$, are given in Table \ref{tab:oxygenP}.
\begin{table}[tb]
\caption{Magnetic polarization of the oxygen sites in
different magnetic states of HoMnO$_3$
(namely, the absolute values of the magnetic moments
at the oxygen sites in $\mu_b$).
The first value was derived from the model analysis for the isolated Mn($3d$) bands, while
the second value (shown in the parentheses) was obtained in the LSDA calculations, which
also
take into account the polarization of the O($2p$) band.
O$_{\bf ab}$ and O$_{\bf c}$ denote the oxygen sites located in the ${\bf ab}$-plane
and between the planes, respectively. Two lines in the case of the
E-phase stand for the polarization in the FM (first line) and AFM (second line)
Mn-O-Mn bonds.
The
finite polarization in some AFM Mn-O-Mn bonds is related
to the oxygen displacements from the midpoint positions in the $D^{16}_{2h}$ structure.}
\label{tab:oxygenP}
\begin{tabular}{ccc}
\hline
 phase   & O$_{\bf ab}$                & O$_{\bf c}$   \\
\hline
 F       &  $0.26~(0.11)$              &  $0.23~(0.04)$  \\
 A       &  $0.25~(0.09)$              &  $0~(0)$     \\
 C       &  $0.08~(0.02)$              &  $0.20~(0.03)$  \\
 G       &  $0.07~(0.03)$              &  $0~(0)$     \\
 E       &  $\begin{array}{c} 0.24~(0.07) \\ 0.09~(0.01) \\ \end{array}$ &  $0~(0)$     \\
\hline
\end{tabular}
\end{table}
The parameters $J_{\rm O}$ can be derived from the LMTO calculations in the
local-spin-density approximation (LSDA).\cite{remark3}
It yields $J_{\rm O}=$ 2.1 and 2.2 eV for the oxygen sites located in the
${\bf ab}$-plane and between the planes, respectively.
Then, the energy gain,
caused by the
polarization of the oxygen sites,
can be estimated from the formula
$\Delta E_{\rm O} = -\frac{1}{4} M_{\rm O}^2$ (with subsequent summation
over all oxygen sites in the formula unit),
which yields
$\Delta E_{\rm O} =$ $-$$102$, $-$$63$, $-$$29$, $-$$5$, and $-$$33$ meV
for the states F, A, C, G, and E, respectively.
The effect is clearly too big. For example, by
by combining these values with the total energies shown in Fig. \ref{fig.TotalE},
we would arrive to the FM ground state, which again
contradicts to the experimental data. Then, what is missing?

  One effect is related to the polarization of the O($2p$)-band,
which is not explicitly included in the model (\ref{eqn:Hmanybody}). It is true that
since the O($2p$)-band is filled, it does not contribute to the total magnetic
moment. However, it can contribute to the local moments, which
cancel each other after the summation over the unit cell.
Particularly, the polarization of the oxygen states in the O($2p$)-band appears to
be the opposite to the one in the Mn($3d$)-band, as it follows from the
form of the Mn($3d$)-O($2p$)
hybridization.\cite{remark6} This effect is clearly seen
by comparing the moments
obtained for the isolated Mn($3d$)-bands with results of the
all-electron calculations, which take
into account the contributions of the O($2p$)-band (Table \ref{tab:oxygenP}). Indeed,
the O($2p$)-band substantially reduces the values of the magnetic moments
associated with
the oxygen sites (by factor two and more).
Therefore, $\Delta E_{\rm O}$ will be also reduced.
For example, by using the LSDA values for $M_{\rm O}$ (Table \ref{tab:oxygenP}), we
find that $\Delta E_{\rm O}$ is reduced till
$-13$, $-9$, $-1$, $-4$, and $-11$ meV per one formula unit
for the states F, A, C, G, and E, respectively.
By combining these $\Delta E_{\rm O}$ with the total energies
shown in Fig. \ref{fig.TotalE},
we readily obtain the E-type AFM structure is realized as the
ground state, in agreement with the experiment.
The new values of the total energies, measured relative to the FM state,
are
$-15$, $-15$, $-27$, and $-33$ meV per one formula unit
for the states A, C, G, and E, respectively.

  Another factor, which strongly affects the relative stability
of different magnetic states, is the correlation interactions
beyond the HF approximation. In order to estimate the energies
of these correlation interactions, we tried three perturbative
techniques starting from the HF solutions for each magnetic state.
One is the random-phase approximation (RPA), which takes into
account the lowest-order polarization processes, involving the
excitation and subsequent deexcitation of an
electron-hole pair.\cite{Pines,BarthHedin,Ferdi02}
For these purposes, the RPA expression for the correlation energy
has been adopted for the model calculations.\cite{remark9}
Another method is the second order perturbation theory for the
correlation interactions,\cite{rev08,PRB06b,JETP07,remark10}
and the third one is the $T$-matrix method,\cite{JETP07,Kanamori3}
which takes into account higher-order effects. Results of these
calculations for HoMnO$_3$ are shown in Table \ref{tab:CorrelationE}.
\begin{table}[tb]
\caption{Correlation energies for several AFM states of HoMnO$_3$ measured
in meV per one formula unit relative to the FM state. The correlation energies
have been computed
in the
random-phase approximation (RPA), the second-order perturbation theory,
and the $T$-matrix method starting from the Hartree-Fock approximation
for each magnetic state.}
\label{tab:CorrelationE}
\begin{tabular}{lcccc}
\hline
 method     & A                & C         & G               & E         \\
\hline
 RPA        &  $-4.9$          &  $-19.5$  &  $-24.7$        &  $-14.1$  \\
 2nd order  &  $-6.7$          &  $-14.9$  &  $-17.9$        &  $-10.3$  \\
 $T$-matrix &  $-4.6$          &  $-9.8$   &  $-11.7$        &  $-7.6$   \\
\hline
\end{tabular}
\end{table}
Since the on-site Coulomb repulsion $U$ is relatively small, all three
methods provide rather consistent explanation for the behavior of the
correlation energies, which tend to stabilize the AFM states relative to the FM one.
The energy gain increases with the number of the AFM bonds in the
direction F$\rightarrow$A$\rightarrow$E$\rightarrow$C$\rightarrow$G.
Thus, the correlation interactions act against the magnetic polarization of the
oxygen sites and again tend to destabilize the E-state relative to the G-state.
The situation is rather fragile and whether the E-state is realized as the
ground state of HoMnO$_3$ depends on the delicate balance of these two
effects and also on the approximations employed for the correlation energy.
For example, RPA and the second-order perturbation theory seem to overestimate the
correlation energy of the G-state and make the E-state unstable.
On the other hand, the E-state, which breaks the orthorhombic
$D_{2h}^{16}$ symmetry, should be additionally stabilized through the
lattice relaxation.

\section{\label{sec:summary}Summary and Conclusions}

  On the basis of first-principles electronic structure calculations,
we propose a microscopic model for the behavior of interatomic
magnetic interactions in the series of orthorhombic manganites
$R$MnO$_3$ ($R$$=$ La, Pr, Nd, Tb, and Ho), which explains the
phase transition from the A-type AFM state to the E-state with
the increase of the lattice distortion.
Our picture
is clearly different from the ones proposed in the previous studies.
In fact, several authors emphasized the importance of the 2nd-neighbor interactions
$J^{\bf a}_2$ and $J^{\bf b}_2$
in the orthorhombic ${\bf ab}$-plane.
For example, Kimura \textit{et al.}\cite{Kimura} considered the superexchange
processes mediated by the O($2p$) orbitals in the distorted perovskite structure
and argues that they can be responsible for the
AFM interaction $J^{\bf b}_2$
and weakly FM interaction $J^{\bf a}_2$. Other authors\cite{Picozzi06,Xiang} performed
the mapping of the total energies derived from the first-principles electronic
structure calculations onto the Heisenberg model and argued that under certain
conditions
$J^{\bf a}_2$ and $J^{\bf b}_2$
become comparable with $J^\parallel_1$. However, such a mapping
crucially depend on the form of the \textit{a priori} postulated model, where
the
lack of some interactions
(such as $J_3$)
can lead to an incomplete picture.
In this sense, our approach to the problem is more consistent.
\begin{itemize}
\item[$\bullet$]
It does not make
any \textit{a priori} assumptions about the form of the Heisenberg model.
\item[$\bullet$]
It goes beyond the conventional superexchange processes
and takes into account other contributions to interatomic magnetic interactions.\cite{TRN}
\end{itemize}
Particularly,
the contributions associated with the ``super-superexchange'' processes
in the regime of relatively small on-site Coulomb interactions,
give rise to the 3rd-neighbor coupling $J_3$, which was overlooked in the
previous studies.\cite{remark4}
According to
our point of view, $J_3$ is one of the key players, which
triggers the transition to the E-type AFM state in
orthorhombic manganites.
\begin{itemize}
\item[$\bullet$]
The existence of $J_3$ is directly related to the form of the
orbital ordering.
\item[$\bullet$]
$J_3$ is responsible for the AFM coupling between 3rd-neighbor
spins in the ${\bf ab}$-plane, which is realized in
the E-phase (Fig. \ref{fig.intro}).
\end{itemize}

  Since the longer range AFM interactions seem to be
the intrinsic property of all undoped manganites, these interactions should
be seen in the experiment, for example, on the inelastic neutron scattering.
We expect the longer range interactions to take place
even in LaMnO$_3$. Although it has A-type AFM ground state,
the longer range interactions participate as the precursors of the E-phase,
which is finally realized in the more distorted compounds.
The neutron-scattering measurements on LaMnO$_3$
are available today. Nevertheless, the
experimental data are typically interpreted only in terms of
the NN interactions.\cite{Hirota,Moussa}
Definitely, the problem deserves further analysis. Particularly, in would be interesting
to reinterpret the experimental data by permitting the longer range interactions,
particularly
$J^{\bf b}_2$ and $J_3$.
This point was already emphasized in ref. \citen{springer}.
It is possible that the
longer range interactions are not particularly strong in LaMnO$_3$, which
has the highest N\'{e}el temperature (${\rm T_N}$, Fig. \ref{fig.intro})
and where the NN interactions clearly dominate.
From this point of view, it would be more interesting to consider two other A-type AFM
systems, PrMnO$_3$ and NdMnO$_3$, which have smaller ${\rm T_N}$ and where the
relative contribution of the
longer range interactions to the magnon spectra is expected to be stronger.

  Although the proposed model is able to unveil the microscopic origin of the
magnetic phase transition, the quantitative agreement with the experimental
data crucially depends on the combination of the following three factors:
\begin{itemize}
\item[$\bullet$]
the correlation effects beyond the HF approximation;
\item[$\bullet$]
the magnetic polarization of the oxygen sites;
\item[$\bullet$]
the lattice relaxation in the E-phase, which breaks the
inversion symmetry and gives rise to the multiferroic behavior.
\end{itemize}
The detailed analysis of these effects presents and interesting and
important problem for the future investigations.

\section*{Acknowledgment}
I am grateful to Zlata Pchelkina for valuable discussions and the
help with preparation of Figs. \ref{fig.OrbitalOrdering}
and \ref{fig.OrbitalOrderingMnOMn}.
The work is partly supported by Grant-in-Aid for Scientific
Research in Priority Area ``Anomalous Quantum Materials''
and Grant-in-Aid for Scientific
Research (C) No. 20540337
from the
Ministry of Education, Culture, Sport, Science and Technology of
Japan.


\begin{thebibliography}{99}
\bibitem{remark7}
Note that according to the formal valence arguments, four $3d$ electrons will reside on the
three $t_{2g}$ levels and the lowest $e_g$ level for the majority-spin states,
while all minority-spin states will be unoccupied.
\bibitem{remark8}
Otherwise, the FM and AFM contributions
to the exchange energy will cancel each other
and the directions of spins can
be arbitrary.
\bibitem{remark2}
More precisely, such a configuration corresponds to the local
maximum of the total energy rather that to a minimum.
\bibitem{remark5}
Note that some aspects of the double exchange physics can be
applied to the band insulators. The situation was discussed
in ref. \citen{BiMnO3}.
\bibitem{remark3}
The intraatomic exchange coupling $J_{\rm O}$ at the oxygen sites was obtained
from the formula $\Delta^{\rm ex}_{\rm O} = J_{\rm O} M_{\rm O}$, where
$\Delta^{\rm ex}_{\rm O}$ is the intraatomic exchange splitting and
$M_{\rm O}$ is the local magnetic moment.
The values of $\Delta^{\rm ex}_{\rm O}$ and $M_{\rm O}$ were taken from the LSDA
calculations for the FM state. Particularly,
$\Delta^{\rm ex}_{\rm O} = C^\downarrow - C^\uparrow$ can be related to the centers
of gravity for the minority- and majority-spin $2p$-states (the so-called
$C$-parameters of the LMTO method -- ref. \citen{LMTO}).
\bibitem{remark6}
Since the $\uparrow$-spin Mn($3d$)-states are closer in energy to the
O($2p$)-states, the former will be strongly admixed into the
$\uparrow$-spin O($2p$)-band in comparison with the similar weight of the
Mn($3d$)-states in the $\downarrow$-spin O($2p$)-band. Since the total
number of electrons in the $\uparrow$- and $\downarrow$-spin
O($2p$)-bands is conserved, the increase/increase of the weight of the
Mn($3d$)-states is compensated by the O($2p$)-states. Therefore, the
O($2p$)-states in the O($2p$)-band
will be polarizaed opposite to the Mn($3d$)-band.
\bibitem{remark10}
Here, the correlation interaction is defined as the difference between
the true many-body Hamiltonian (\ref{eqn:Hmanybody}) and its
one-electron counterpart, obtained at the level of the
HF approximation.
\end{thebibliography}
\end{document}